%% file: main.tex
\documentclass[conference]{IEEEtran}
\IEEEoverridecommandlockouts
\usepackage{cite}
\usepackage{amsmath,amssymb,amsfonts}
\usepackage{algorithmic}
\usepackage{graphicx}
\usepackage{textcomp}
\usepackage{xcolor}

\usepackage{enumerate}
\usepackage{multirow}
\usepackage{times}
\usepackage[colorinlistoftodos]{todonotes}
\usepackage{url}
\usepackage{enumitem}
\usepackage{listings}
\usepackage{subcaption}
\usepackage{caption}
\usepackage{balance}

\def\BibTeX{{\rm B\kern-.05em{\sc i\kern-.025em b}\kern-.08em
    T\kern-.1667em\lower.7ex\hbox{E}\kern-.125emX}}

\begin{document}

\title{SensiX: A Platform for Collaborative Machine Learning on the Edge}

\makeatletter
\newcommand{\linebreakand}{%
  \end{@IEEEauthorhalign}
  \hfill\mbox{}\par
  \mbox{}\hfill\begin{@IEEEauthorhalign}
}
\makeatother

\author{\IEEEauthorblockN{Chulhong Min}
\IEEEauthorblockA{\textit{Nokia Bell Labs} \\
Cambridge, UK \\
chulhong.min@nokia-bell-labs.com}
\and
\IEEEauthorblockN{Akhil Mathur}
\IEEEauthorblockA{\textit{Nokia Bell Labs} \\
Cambridge, UK \\
akhil.mathur@nokia-bell-labs.com}
\and
\IEEEauthorblockN{Alessandro Montanari}
\IEEEauthorblockA{\textit{Nokia Bell Labs} \\
Cambridge, UK \\
alessandro.montanari@nokia-bell-labs.com }
\linebreakand
\IEEEauthorblockN{Utku G\"{u}nay Acer}
\IEEEauthorblockA{\textit{Nokia Bell Labs} \\
Antwerp, Belgium \\
utku\_gunay.acer@nokia-bell-labs.com}
\and
\IEEEauthorblockN{Fahim Kawsar}
\IEEEauthorblockA{\textit{Nokia Bell Labs} \\
Cambridge, UK \\
fahim.kawsar@nokia-bell-labs.com}
}

\maketitle

\input{sections/abstract}


\input{sections/introduction.tex}

\input{sections/background.tex}
\input{sections/design.tex}
\input{sections/system.tex}

\input{sections/evaluation.tex}

\input{sections/discussion.tex}
\input{sections/related_work.tex}

\input{sections/conclusion.tex}

\bibliographystyle{ieeetr}
\bibliography{reference.bib}

\balance

\end{document}

%% file: sections/abstract.tex
\begin{abstract}

The emergence of multiple sensory devices on or near a human body is uncovering a new dynamics of extreme edge computing. In this, a powerful and resource-rich edge device such as a smartphone or a Wi-Fi gateway is transformed into a \textit{Personal Edge}, collaborating with multiple devices to offer remarkable sensory applications, while harnessing the power of locality, availability and proximity. Naturally, this transformation is pushing us to rethink on how to construct accurate, robust, and efficient sensory systems in this personal edge environment. For instance, how do we build a reliable activity tracker with multiple on-body devices equipped with IMUs? While the accuracy of sensing models are improving, their runtime performance still suffers, especially under this emerging multi-device, personal edge environments. Two prime caveats that impact their performance are device and data variabilities, contributed by several runtime factors including device availability, sensing hardware, data quality, resource budget, and device placement. To this end, we present SensiX, a system component for the personal edge, that stays between sensor data and sensing models and ensures \textit{best-effort inference} quality under any condition while coping with device and data variabilities without demanding model engineering. SensiX externalises models execution away from the application, and comprise of two essential functions, a neural translation operator for principled mapping of device-to-device data and a quality-aware selection operator to systematically choose the right execution path as a function of model accuracy.  Collectively, these operators automate the sensing model execution in a multi-device, personal edge environment. We report the design and implementation of SensiX and demonstrate its efficacy in developing motion and audio-based multi-device sensing systems with a personal edge. Our evaluation shows that SensiX offers 7-13\% increase in overall accuracy and up to 30\% increase across different environment dynamics at the expense of 3mW power overhead.

\end{abstract}




%% file: sections/introduction.tex
\section{Introduction}

Edge AI is coming to our lives. Recent AI accelerators\footnote{Here, AI accelerators refer to a class of brand-new purpose-built System On a Chip (SoC) for running deep learning models efficiently on edge devices.} bring AI to personal devices ranging from smartphones (e.g., Pixel Neural Core on Pixel 4, A13 Bionic chip on iPhone 11) to embedded devices (e.g., Nvidia Jetson Nano~\cite{nvidia_jetson_nano}, Google Coral~\cite{google_coral}, Intel Neural Compute Stick 2~\cite{intel_ncs2}). They are enabling powerful, cloud-scale AI to run anytime and anywhere solely on personal computing environments, without relying on clouds and even cloudlets. Sensory devices are also now pervasive and surrounding us. These mobile, wearable, and IoT devices on and near our body are increasingly embracing AI/ML algorithms to uncover remarkable sensory applications~\cite{miluzzo2008sensing,lane2010survey,mohan2008nericell,kang2008seemon,rachuri2010emotionsense,liang2019audio, yao2018qualitydeepsense,kawsar2018earables}. In this transformation, we are observing the emergence of \emph{personal edge} as a natural course of AI accelerators and multiple sensory devices surrounding us. A resource-rich personal edge device such as a smartphone, or a nearby Wi-Fi gateway is now transformed into an extreme edge node and collaborates with multiple sensory devices to build powerful sensory application while leveraging the benefits of locality, availability and proximity.

\begin{figure}
    \centering
    \includegraphics[width=1.0\columnwidth]{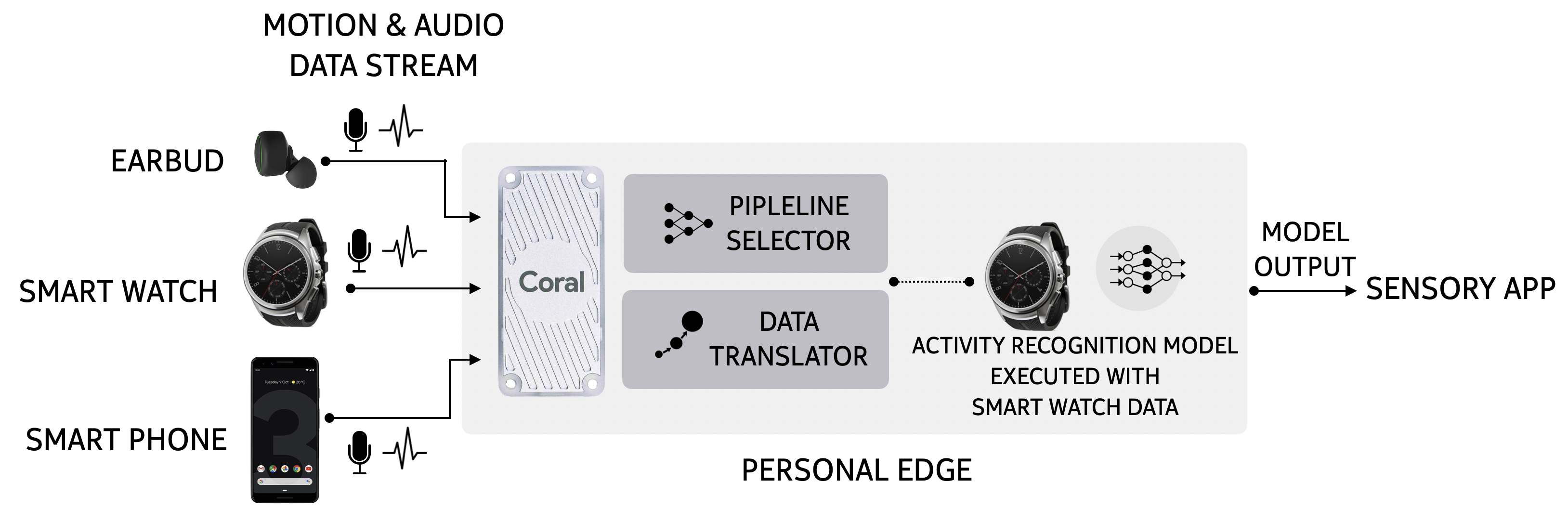}
   \caption{SensiX stays between sensory devices and sensory model(s) in a personal edge and applies neural translation and neural selection operators to selectively chose the execution path for best-effort inference. Here, SensiX uses smartwatch data and respective model execution path to guarantee the best accuracy based on the outcome of translation, and selection operators. }
    \label{fig:sensix_overview}
\end{figure}

Beyond the AI capability itself, the multiplicity of sensory devices is further opening up an exciting opportunity to leverage sensor redundancy and high availability afforded by multiple devices. For instance, a personal health tracker or a cognitive assistance application can now selectively use a smartwatch, a smartphone, or a smart earbud for the sensor data to ensure the robust and accurate performance of the model and the application. However, such an advantage comes at the expense of increasing complexity. Two key caveats that contribute to this complexity are \emph{device and data variabilities}, caused by runtime factors including device availability, sensing hardware~\cite{stisen2015smart}, resource budget, data quality and device placement~\cite{min2019early}. For instance, in the examples above, each of the three devices offers identical sensing modalities (e.g., motion and audio), but with different sensing quality due to runtime behaviour. Thus, careless selection of devices and naive use of sensing models would cause unexpected degradation of runtime accuracy. To optimise the accuracy of sensing models, application developers often designed a purpose-built sensing pipeline optimised for a specific device or a specific environment. However, such tight coupling naturally suffers in multi-device environments because these runtime variabilities are incredibly hard to anticipate during the model design and training time. Thus, it is imperative to build a sensing system in this multi-device personal edge environment that takes these attributes into account at runtime and ensure the best \emph{runtime accuracy} under any given conditions.

There have been extensive studies on the multiplicity of sensors in the past, and these works contributed substantially to advance our understanding of multi-device sensing research. In body sensor network literature, there have been attempts to build runtime frameworks with on-body sensors for context awareness~\cite{zappi2008activity,kang2008seemon,kang2010orchestrator,keally2011pbn}. These work mainly promote dynamic device discovery and selection for resource management and seamless operation with a static view on average accuracy. In multi-sensory fusion research, several studies looked at model optimisation strategies while addressing system issues such as time synchronisation and missing data~\cite{ordonez2016deep,peng2018aroma,yao2017deepsense,yao2018qualitydeepsense,vaizman2018context}.  
However, due to the tight coupling of the device combinations, these methods are neither practical nor scalable in this highly fluid and dynamic multi-device environments. It would be incredibly hard, if not impossible, to train and deploy different fusion models for all possible combinations of devices.

In contrast, in this work, we posit the question: \emph{When sensing models are given, how can we address the challenges of device and data variabilities at runtime to ensure best-effort inference quality under any condition in a multi-device environment?"}

We define \textit{best-effort inference} as the optimum model accuracy guaranteed by the system given multiple execution choices. To this end, we report the design and development of SensiX, a brand-new system component for personal edge offering best-effort inference in a multi-device sensing system. SensiX stays between different sensory devices and corresponding model(s) in a personal edge environment and performs principled data engineering to select the best execution path as a function of model accuracy while externalising model management and execution away from the application. SensiX achieves this with two purpose-built neural operations, as shown in Figure \ref{fig:sensix_overview}. First, a neural translation operator that deals with device variability by mapping data across devices, and second, a runtime quality assessment operator that deals with data variability by selecting the right execution path for the best model accuracy. Collectively, these two operators enable SensiX to dynamically and automatically compose a model execution path under any condition to ensure best-effort inference while coping with runtime device and data heterogeneity.

We evaluate SensiX on two representative multi-device sensing applications built with motion and audio signals for physical activity and keyword recognition.  Our results suggest that SensiX offers 7-13\% increase in overall accuracy and up to 30\% increase across different environment dynamics. This performance gain comes at the expense of 3mW on the host device, however with a significant reduction of development complexity and cost.

In what follows, we discuss the unique characteristics of multi-device sensing systems and present corresponding challenges of multi-device sensing systems to inform our design decisions. Next, we describe the technical details of SensiX and its different operations. We then move to the evaluation of SensiX and reflect on some critical issues before concluding the paper.

%% file: sections/background.tex
\section{Background}

Over several decades, extensive studies on machine learning and artificial intelligence have been actively conducted to understand us and the world around us from raw sensory signals. Since these algorithms, especially deep learning-based ones, naturally require a substantial amount of computation, much effort has also been put toward enabling them to run on resource-constraint devices, e.g., using model offloading and partitioning~\cite{cuervo2010maui,ha2014towards,bahreini2017efficient}, model compression~\cite{lane2016deepx,liu2018demand,reagen2016minerva}. On the basis of the insights from these works, mobile runtime systems have been studied and built, which take inference pipelines (or sensing model) as the main workload and manage their performance over dynamic, heterogeneous environments, especially in terms of accuracy, energy consumption, and latency. In common, they exploit alternative, substitutable processing options for given inference pipelines, e.g., at the level of sensor~\cite{kang2010orchestrator}, 
processor~\cite{georgiev2016leo}, hyper-parameter of classifiers~\cite{chu2011balancing}, approximation of model~\cite{han2016mcdnn}, and dynamically select the best one based on their expected quality and the system policy.

While the quality (and corresponding performance) of a sensing model is dynamic, the runtime assessment of those systems is mostly limited to resource metrics such as energy cost and latency. On the contrary, they relatively have a static view on the accuracy of sensing models and often relied on their average accuracy, obtained in the training phase. It worked in the conventional, single-device environment where the fixed, same device is supposed to be used. However, emerging multi-device environments bring new challenges, \emph{device and data variabilities} that make runtime accuracy of sensing models dynamic and unpredictable. We present data-driven evidence and reflect on key design challenges of the system for ensuring best-effort inference quality in a multi-device environment.

\begin{figure}
    \centering
    \includegraphics[width=1.0\columnwidth]{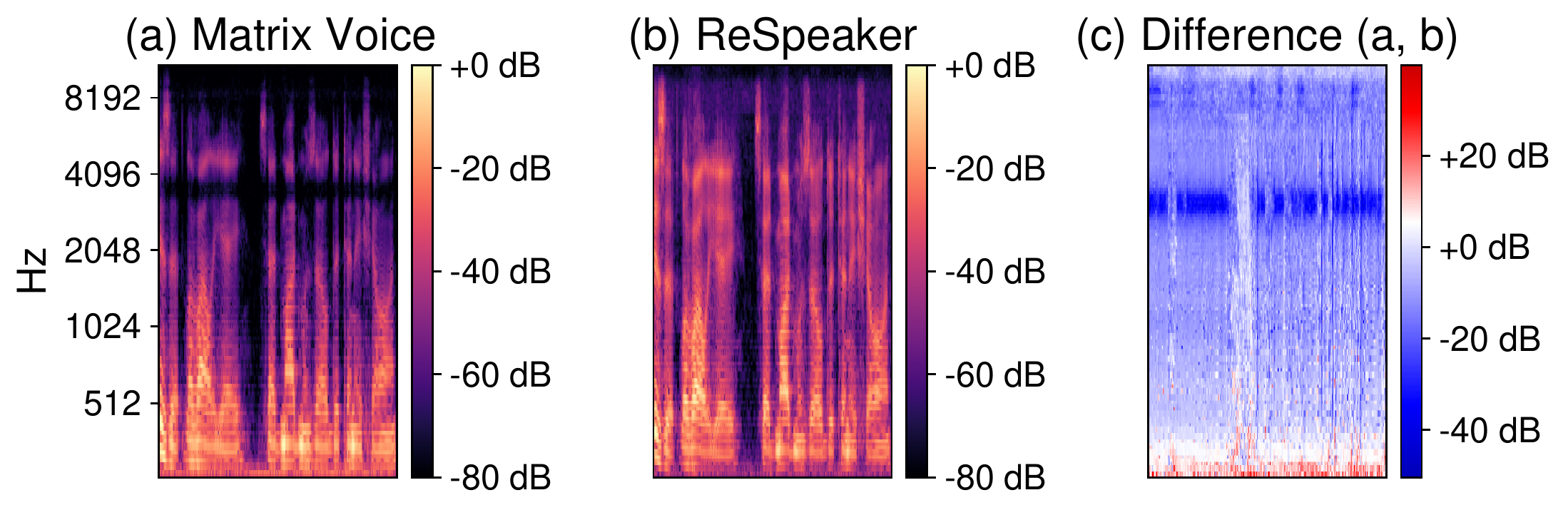}
   \caption{Mel-spectrograms of a speech segment as captured by (a) Matrix Voice, (b) ReSpeaker, and (c) their difference.}
    \label{fig:data_uniformity}
\end{figure}

\textbf{Device variability: } Even before a sensor signal reaches the sensory application (e.g., a classifier), it passes through several processing stages including ADC conversion, DSP processing, OS processing -- each of which can introduce some artefacts in the signal. Naturally, these artefacts vary across devices, as such different devices capture the same physical phenomena slightly differently. This \emph{heterogeneity} characteristic has a profound impact on model performance, especially when different devices are available for data acquisition. To provide empirical evidence, in Figure~\ref{fig:data_uniformity}, we show mel-spectrograms of a 3-second speech segment as recorded by the two microphones (Matrix Voice and ReSpeaker) simultaneously. We observe that the microphones exhibit differences in their frequency responses to the same speech input - also visualised in the rightmost figure. In~\cite{stisen2015smart}, the impact of heterogeneous IMU sensor on the activity recognition performance has also been thoroughly studied. Since these variations are common characteristics when different, heterogeneous devices are involved in a sensing task irrespective of the modalities, unexpected performance degradation would be inevitable if a pre-trained model is deployed in unseen devices or shared with different devices. A straightforward solution would be to train device-specific models, but it would be almost infeasible considering device heterogeneity in today's market.

\begin{figure}
    \centering
    \includegraphics[width=0.9\columnwidth]{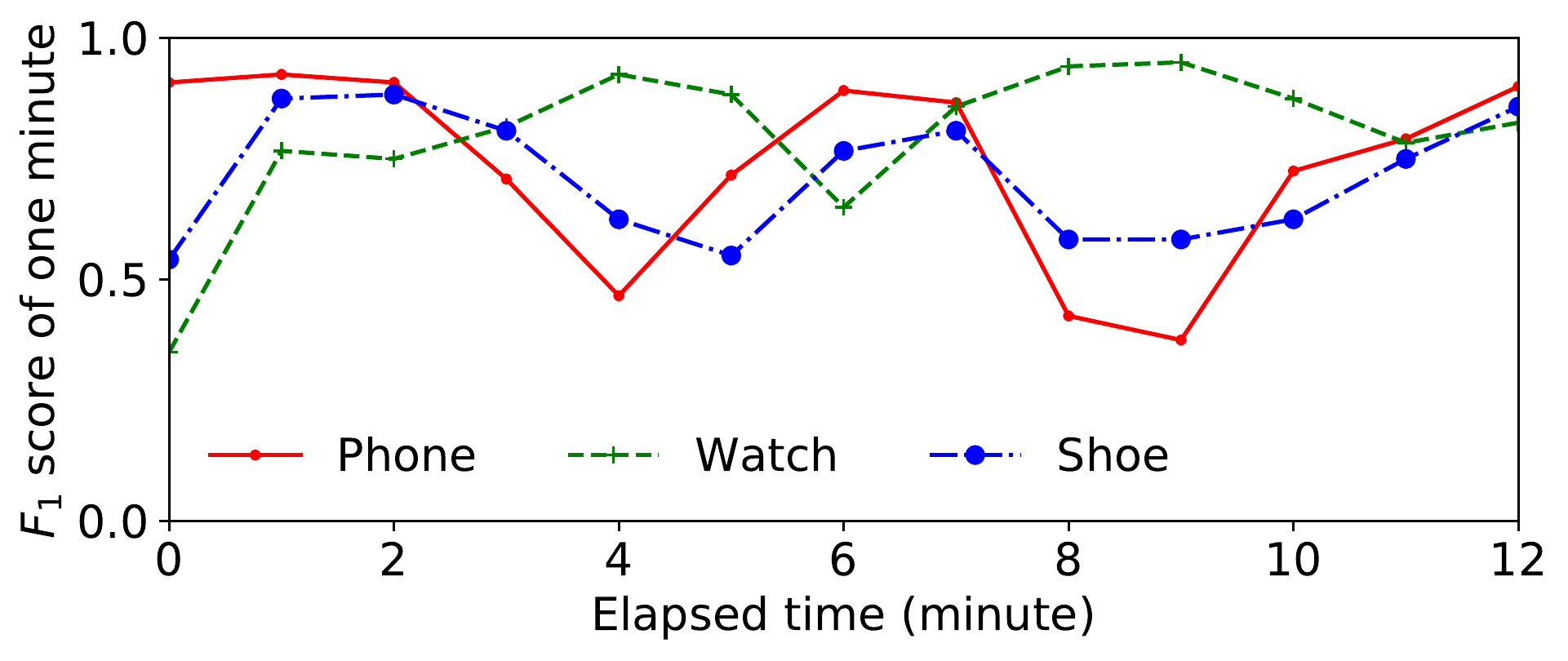}
   \caption{Runtime accuracy of activity recognition model~\cite{peng2018aroma} trained with the Opportunity dataset~\cite{roggen2010collecting}; 1) The same device offers varying accuracy over time and 2) different devices offer the best accuracy at different times}
    \label{fig:data_quality}
\end{figure}

\textbf{Data variability: } Many sensory devices around us share a standard set of sensors, e.g., IMU, microphone, etc., offering \emph{redundant, substitutable} sensing capabilities. For example, a keyword spotting model can selectively run on one of the available microphones around a user. Similarly, an activity recognition model can be performed with one of the IMU-equipped wearables, e.g., a smartphone, a smartwatch, or even an earbud. There are several factors that constitute and affect the expected runtime accuracy, spanning hardware~\cite{stisen2015smart}, device placement, and even users' behavioural characteristics~\cite{kreil2014dealing, min2019early}. The dynamic nature of these factors makes runtime accuracy dynamic even with the fixed composition of an inference pipeline (i.e., sensor stream from the same device with the same sensing model.) To quantify this aspect, we provide empirical evidence with a motion model in Figure~\ref{fig:data_quality}. With the Opportunity dataset~\cite{roggen2010collecting}, we selected three IMU devices placed on a hip, a left lower arm, and a right shoe, which can be mapped to the typical position of a smartphone, a smartwatch and a smart shoe, respectively. Then, we trained three activity recognition models~\cite{peng2018aroma} separately for each device and observed how $F_1$ score of these models changes every minute. The results show that 1) each device offers varying accuracy over time and 2) more importantly, the best performing device also changes over time.

The goodness of a sensing model often determines the goodness of a sensing system. However, while the above challenges do not necessarily contribute to the goodness of a sensing model at a training time (which is solely dependent on the quality, quantity, and diversity of training data, training strategy, and model architecture), it is important to note that the combined effect of these factors could significantly degrade the runtime performance of the sensing model.


%% file: sections/design.tex
\section{SensiX Design}

\subsection{Design Goal}

When a sensing model is deployed and running in unseen situations, accuracy degradation to some extent is inevitable due to \emph{device and data variabilities}, especially compared to the accuracy obtained in the training phase. Recently, extensive studies have been conducted to compensate such a gap, e.g., by adopting data augmentation~\cite{ho2019population,um2017data}, transfer learning, and incremental learning. They show remarkable performance improvement without training an entirely new model, but still require an additional burden for data collection or model engineering.

In this paper, we propose a novel approach of boosting the runtime accuracy of sensing models, i.e., the \emph{system-driven best-effort inference}. Without relying on model engineering and application modification, it actively intervenes between sensors and sensing models, and achieves the accuracy improvement by dynamically addressing device and data variabilities in an autonomous manner.


\begin{figure*}[t]
    \centering
    \includegraphics[width=0.7\textwidth]{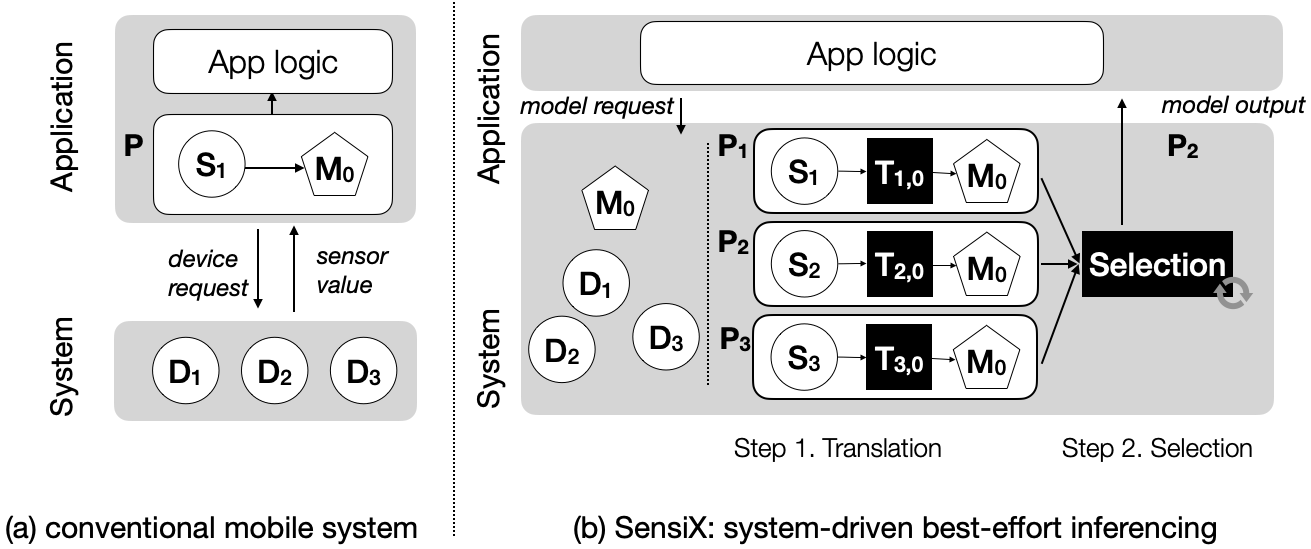}
    \caption{Overview of SensiX operation; D, S, M, and T represents devices, sensor data, models and translation functions, respectively.}
    \label{fig:overview}
\end{figure*}

\subsection{Separating Execution of Sensing Models from Application}

In conventional sensing systems, applications are entirely responsible for the execution and management of sensing models. As shown in Figure~\ref{fig:overview} (a), an application requests sensor data of interest (e.g., IMU data or audio data) to the underlying system and manages the end-to-end operations required for model processing, in the application space, spanning over data collection, model design and tuning, deployment, etc. While there are public model hubs available for pre-trained models, e.g., Tensorflow Hub~\cite{tensorflow_hub}, PyTorch Hub~\cite{pytorch_hub}, and ModelHub~\cite{hosny2019modelhubai}, it is still mostly developers' burden to construct the full execution pipeline from raw sensor data.

A critical aspect of \emph{system-driven} best-effort inference is to separate the execution complexities of a sensing model from the model training process as well as from the application logic (Figure~\ref{fig:overview} (b)). This facet is particularly essential in several aspects. First, it can take significant, but duplicate efforts away from application developers and model experts. As the execution of sensing models becomes a core, increasingly demanding operation in sensing environments, such separation naturally becomes a key requirement of sensing systems; similar to how traditional OSes have abstracted basic tasks from computer programs, such as handling I/O, controlling peripheral devices, etc. Second, it is almost infeasible for application developers and model experts to address those \emph{runtime} properties in advance at the development time and model training time. On the contrary, the system can intervene with the execution of sensing models actively and dynamically as it has more visibility and fine-grained control over device and data variabilities. 

\subsubsection{Level of abstraction}

Separating execution operations from application space demands articulated and useful abstraction hiding the underlying complexities. One of the design challenges is to determine at which level of sensing pipelines is abstracted. A typical pipeline of sensing models consists of sensing, preprocessing, running models, and delivering context outputs. Accordingly, we can imagine the level of abstraction corresponding to each level of operation. In this paper, we adopt an abstract at the \textit{sensing model} scale. That is, applications specify sensing models of interest for their logic, and SensiX takes care of the full operations required to execute the models. Our design decision in Figure~\ref{fig:overview} (b) offers several benefits, including application-side flexibility of choosing sensing models, a significant reduction of code complexity, and better system-wide management and coordination.

Our design choice is different from modern mobile operating systems in diverse aspects. For example, current sensor libraries, e.g., \texttt{SensorManager}~\footnote{\url{https://developer.android.com/reference/android/hardware/SensorManager}} on Android, can be seen to provide an abstraction of \textit{sensing}. That is, the system handles hardware operation for sensor readings, and the application takes care of the rest of the pipelines (See Figure~\ref{fig:overview} (a)). In this case, applications have a high level of flexibility, but it gives significant burdens to developers at the same time. The system also has very little room for performance optimisation. In another extreme, Android also provides the abstraction as a unit of \textit{high-level context}, e.g., \textit{Activity Recognition API}~\footnote{\url{https://developers.google.com/location-context/activity-recognition}} on Android. If an application specifies the context type of interest, e.g., \textit{activity}, the system takes care of the full sensing pipeline for activity recognition and delivers the final result only. It relieves all the burdens for context inference from application developers and gives the system more flexibility, e.g., choice and scheduling of sensing models, for the system-wide optimisation of the resource use. However, it limits the flexibility of application logic and requires a very well-defined and pre-established taxonomy of context vocabulary. 

\subsection{Dynamic Translation and Selection}

To dynamically address device and data variabilities at runtime in multi-device environments, we devise two purpose-built neural operators that actively act in the middle of sensors and sensing models, \emph{device-to-device data translator} (\S\ref{subsec:sensix_translation}) and \emph{quality-aware runtime pipeline selector} (\S\ref{subsec:sensix_selection}). First, the device-to-device data translator minimises the data variation across devices using a machine-learned component. When the model is deployed in unseen, different devices, SensiX collects unlabelled data in the background and learns the translation function with the collected data, which maps sensor data from a new device to its equivalent point in the training distribution. Second, the quality-aware runtime pipeline selector estimates the data quality of available pipelines (a pair of translated signals and sensing model) at runtime and selects a pipeline offering the best quality.

Figure~\ref{fig:overview} (b) shows the example operation of SensiX when three devices ($D_1$, $D_2$, $D_3$) are available and one sensing model ($M_0$) trained with $D_0$ is given. SensiX first creates the device-to-device data translation functions ($T_{1,0}$, $T_{2,0}$, $T_{3,0}$) for all devices and generates multiple, substitutable pipelines. For example, $T_{1,0}$ makes the data from the sensor $S_1$ on $D_1$ similar to the data from $D_0$ and the corresponding pipeline is generated by composing a sensor (device), a translation function, and a model, e.g., $D_1$, $T_{1,0}$, and $M_0$. Then, SensiX periodically assesses the data qualify of each pipeline and then dynamically selects the best one. We put the translation operator prior to the selection operator because the translation affects the data quality.

%% file: sections/system.tex
\section{SensiX Operation}

\subsection{Device-to-Device Data Translation}~\label{subsec:sensix_translation}

Prior literature~\cite{Das:2014:YHI:2660267.2660325, AccelPrint14, blunck2013heterogeneity} on sensing systems has established that heterogeneities in sensor data are omnipresent and can be caused by a number of issues including variability in hardware, software or usage dynamics of the sensing devices. More critically, it has been shown that even subtle variabilities in sensor data can potentially degrade the performance of state-of-the-art sensing models~\cite{blunck2013heterogeneity, mathur2019mic2mic, mathur2018using}. Indeed, this poses a major challenge for multi-device sensing systems, wherein there is a very high likelihood of variability in the devices owned by the user. For instance, a user may have multiple microphone-enabled devices on (e.g., an Apple iPhone) or near their body (e.g., Amazon Echo), each of which can capture the user's speech and process it through a speech recognition model to understand the user's intent. However, due to the variations in microphone hardware and software processing pipelines across manufacturers, a model trained on an Apple iPhone microphone may not work as well for an Amazon Echo. 



A simple solution to this problem is to train a separate model for each device in the system, however this would incur significant costs to collect and \emph{label} training data for each new device that is added to the system. Instead, SensiX makes a practical and scalable choice: it assumes that there is one sensing model that is shared across all the devices — this model could be trained by a developer either on data collected from one of the devices or even on a separate training dataset independent of the devices owned by the user. Further, we assume that the developer does not provide access to the weights of this pre-trained sensing model, i.e., it is a black-box model — this enables SensiX to support both open-source as well as proprietary models.  

Under these practical assumptions, the technical challenge is to enable highly accurate sensing on multiple heterogeneous devices, even when the sensing model may not have been trained on the same device. To this end, SensiX provides support for \emph{device-to-device data translation} which maps (or translates) a given sensor data from any device to its equivalent point in the training distribution. This translation happens transparently at inference-time (or test-time) and aims at reducing the discrepancy between the test data and the training data on which the sensing model was trained. Overall, the device-to-device translation operation aims to boost the accuracy of the pre-trained sensing model on those devices whose data distribution might differ from the training distribution. Note that when the training and test device are the same, the translation operation is not needed and is ignored by SensiX.


The translation component is machine-learned and based on the principles of Cyclic Generative Adversarial Networks (CycleGAN) as proposed in \cite{mathur2019mic2mic, zhu2017unpaired}. We extend this prior work to support translation of various data modalities, including IMU data (an accelerometer and a gyroscope) and audio data from a microphone. 

\begin{figure}[t]
    \centering
    \includegraphics[width=1.0\columnwidth]{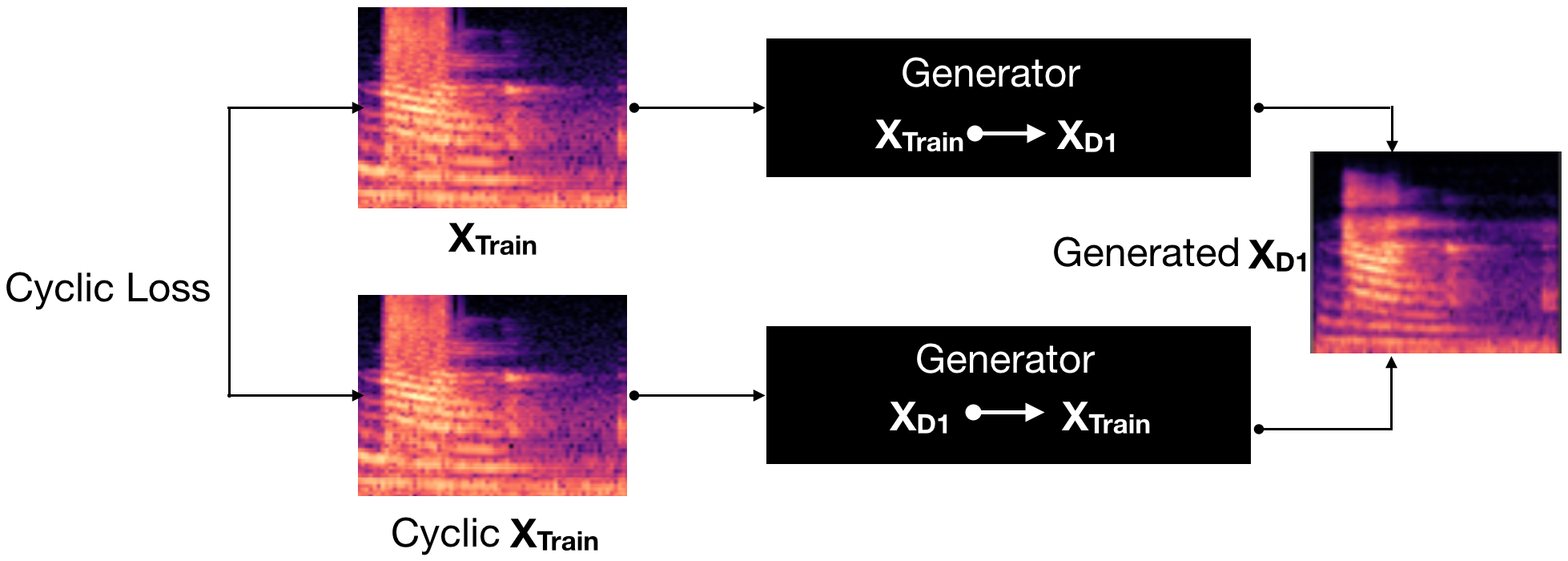}
    \caption{Translation process for $X_{D_{1}} \rightarrow X_{Train}$}
    \label{fig:translation}
\end{figure}

We learn a pair-wise translation function between each user device and the training device (on which the sensing model was trained). Prior works have shown that CycleGAN models can learn these mapping between data distributions solely based on unlabelled and unpaired data, which significantly reduces the cost of training the translation model. We assume that at the time of releasing the sensing model, its developer also provides a small amount of unlabelled data $X_{Train}$ sampled from the training distribution and it is stored on the host device. Further, when a new device $D_1$ is added to the multi-device ecosystem, SensiX collects a small amount of unlabelled data $X_{D_{1}}$ from it with a user's permission and also sends it to the host device. This data need not be time-aligned or paired with the training data $X_{Train}$. Upon receiving unlabelled datasets $X_{Train}$ and $X_{D_{1}}$, the host device initiates the training of a translation mapping $X_{D_{1}} \rightarrow X_{Train}$ based on the CycleGAN architecture as shown in Figure~\ref{fig:translation}. The CycleGAN architecture consists of four neural networks (2 generators and 2 discriminators) that are jointly optimised using adversarial learning -- in our implementation, we use a 6-layer CNN with residual blocks to train the generators and a 4-layer CNN to train the discriminators. Note that such a training operation is conducted only once. Once the training process is done, the trained generator $X_{D_{1}} \rightarrow X_{Train}$ is used to perform real-time translation of the sensor data collected from device $D_1$ to make it similar to the training data. After the translation, it is then passed to the next operations of SensiX for further processing and computing the inferences.  






\subsection{Quality-Aware Pipeline Selection}~\label{subsec:sensix_selection}

After obtaining the translation function for each device, SensiX constructs execution pipelines for each available device. Then, the next question is how to choose a right pipeline out of multiple candidates. We identify two challenges that have to be addressed to make the system practical; (1) how to quantify the quality and (2) how to minimise the system cost while maximising the selection benefit?

\textbf{Quality quantification: } There have been prior attempts to quantify the quality of sensory signals in the domain of signal processing. The most representative example is signal-to-noise ratio (SNR) which was proposed to assess the purity of a signal. However, such signal-level quality assessment is not suitable to assess the quality of an execution pipeline, i.e., its expected runtime accuracy, because the subsequent operations (translation function and sensing model) also affect the quality of the model output. For example, the motion signals on a smartwatch in walking situations would be considered to have the high quality for movement detection, but can be seen to have the low quality for hand gesture recognition.

To address this, we present a novel pipeline-level quality assessment by adopting and modifying the heuristic-based quality assessment (HQA) method proposed in~\cite{min2019closer}. Its key idea is to leverage confidence values reported from a classifier in the sensing model for given (translated) sensory signals and quantify the quality of the execution pipeline based on the values. Confidence values represent how confident a classifier is on the inference output from a given input data. For example, a final softmax layer in neural networks produces a list of probabilities of given sensor data being a member of each class. Leveraging such characteristics, the concept of \emph{uncertainty} of inference output was proposed in the domain of active learning, which represents how uncertain a given inference instance is to be labelled. Inspired by this, we quantify the quality of an execution pipeline by adopting margin sampling~\cite{scheffer2001active}. Margin sampling is computed by taking the difference between the probabilities of the two most likely classes. We can consider that an inference instance with higher margin is more certain to be labelled compare to the other instance with lower margin.

\begin{figure}[t]
    \includegraphics[width=1.0\columnwidth]{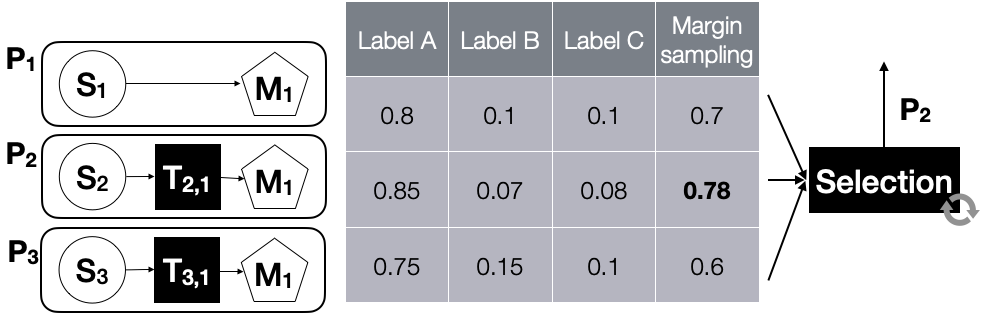}
    \caption{Quality-aware Selection; (S)ensor, (T)ranslation, (M)odel. This is an example when three devices are availale and the mode trained one of these deivces is given.}
    \label{fig:selection}
\end{figure}

Figure~\ref{fig:selection} shows the operational flow and example of the quality-aware selection. SensiX gathers the sensor data from all available devices and obtains the translated data (if the device is different from the device used for training the input model). SensiX executes the model inference with the (translated) data and obtains confidence values for all pipeline candidates. Then, for each pipeline, SensiX computes the margin sampling, i.e., the difference of probabilities between its first and second most probable labels and selects the execution pipeline which shows the highest margin sampling.

\textbf{Energy-efficient selection: } A practical issue in runtime selection is to determine a proper interval of the selection. Since the sensing quality dynamically changes even with the same topology of devices, continuous quality assessment is needed. However, our quality assessment requires all execution pipelines to be performed, i.e., sensing, translating, and model execution, thus the frequent selection could incur a significant system overhead in terms of energy and CPU. To avoid such cost and make the system practical, we adopt a widely used duty-cycling technique for the quality-aware selection. That is, by leveraging temporal locality of human/device contexts, SensiX performs the quality assessment and selection periodically at the fixed interval and deactivates unselected pipelines until the next interval. In this paper, we set the selection interval to 10 seconds. It is important to note that the optimal interval can be different depending on the inference model and user situations and the variable interval is also possible with the prediction of future situations. We leave them as future work. Besides the periodic interval, SensiX also triggers the selection immediately if it detects system events that can affect the selection, e.g., registration and reregistration of a model, and join and leave of a sensor device.

\subsection{SensiX Prototype}~\label{subsec:implementation}

\begin{figure}[t]
    \centering
    \includegraphics[width=0.9\columnwidth]{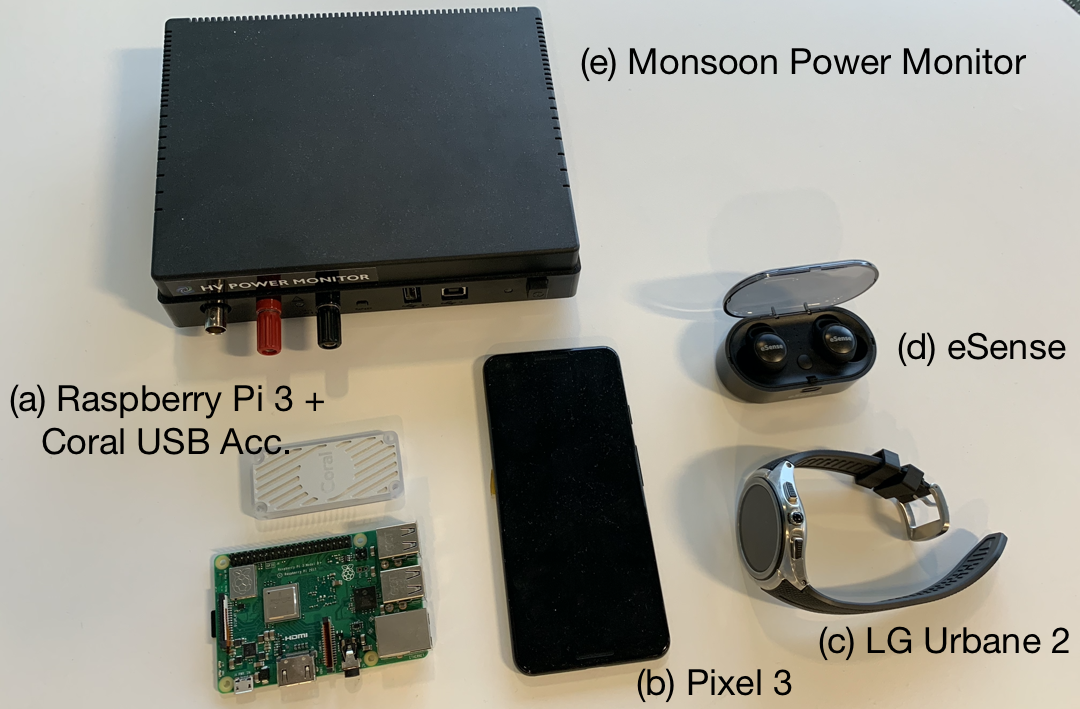}
    \caption{Hardware setup: personal edge as (a) Raspberry Pi 3 with Google Coral USB accelerator, sensor devices as (b) Pixel 3 smartphone, (c) LG Urbane 2 smartwatch, (d) eSense earbuds, and (e) Monsoon power monitor}
    \label{fig:hardware}
\end{figure}

We implement the SensiX prototype on the off-the-shelf devices. Figure~\ref{fig:hardware} shows the hardware setup. For the host device, we used Raspberry Pi 3 with Google Coral USB accelerator and developed the host device-side components with Python. For the sensor devices, we considered three devices, Pixel 3 smartphone, LG Urbane 2 smartwatch, and eSense~\cite{kawsar2018earables}) earbuds. For the Android devices, we developed the sensor-side components that run as an Android background service. For eSense, it does not have processing capability on the board and thus we developed an eSense broker with Python which runs on the host device.

A multi-device sensing system that forms a sensing device pool dynamically and opportunistically for model execution naturally requires a \emph{host} that orchestrates the runtime operations of the system. In a conventional system, a resource-rich device, e.g., a smartphone in personal sensing environments, or a powered Wi-Fi gateway in building environments, etc. can be assigned with such functionalities. The operations discussed earlier, some of which are neural operations demands careful system-wide orchestration and needs to be executed actively in the background. However, modern smartphones are optimised exclusively for maximising battery life, as the OSes often ignore any background operation that has relatively high energy expenses. A clear indication of such decisions of modern mobile OSes is the constrained imposed of background processing for applications. These restrictions have severe implications for the performance of a runtime orchestrator as system-wide optimisation opportunities are relatively limited. 

With a proliferation of multi-device sensing systems, future OSes may remove such restrictions. We consider runtime orchestration and related system operations for multi-device sensing should be designed as a specialised subsystem. Deep learning algorithms drive many sensory models today, and it is natural to expect neural accelerators will power these subsystems, which are not adopted yet to current smartphones. To this end, in SensiX, we have taken a dedicated subsystem route for hosting the runtime orchestrator and implemented it on top of an AI accelerator. However, we envision that SensiX can co-exist with future smartphones (or another prominent mobile form).

This centralised orchestration and execution mean that SensiX considers every sensor devices in the environment as merely a sensor stream provider equipped with a supported communication interface. However, if sensor devices afford processing capability, resource-related system cost would be expected to reduce, e.g., by using code offloading to sensors or pipeline partitioning~\cite{cuervo2010maui, newton2009wishbone}. 

%% file: sections/evaluation.tex
\section{Evaluation}~\label{sec:evaluation}



We present extensive experiments to evaluate the effectiveness of SensiX in multi-device scenarios. We use two multi-device sensing applications built with motion and audio signals for physical activity and keyword spotting, respectively. First, we investigate the effect of our device-to-device data translation and quality-aware selection mechanisms on the runtime accuracy and robustness of sensing models using the multi-device datasets. Then, we perform micro-benchmarks on top of our system prototype to understand its overheads of the SensiX operations.


\subsection{Experimental Setup}

\textbf{Sensing tasks, models, and datasets:} For the evaluation, we use two sensing tasks: \textit{human activity recognition} (HAR) with IMU data and \textit{keyword spotting} with audio data. We choose these two tasks because IMUs and microphones are core sensors in personal-edge, multi-device environments and HAR and keyword spotting tasks are representative ones for these sensors. For comprehensive analysis, we used the mutli-device datasets and conducted repetitive experiments with different combinations of system parameters and comparison groups.

\textit{HAR:} For human activity recognition, we develop a deep learning model proposed in~\cite{peng2018aroma}, which employs a CNN-based feature extractor with 4 residual blocks containing 2 convolutional layers each; it takes 1-second-long data as an input. The model has two fully-connected layers and an output layer. For the analysis, we use the \emph{RealWorld} dataset~\cite{sztyler2016body}. It consists of sensor data recorded from 15 participants with seven smartphones on their body; each participant performs eight physical activities; walking, running, sitting, standing, lying, stairs up, stairs down, and jumping. For the experiments, we select 3-axis accelerometer and 3-gyroscope data from three devices deployed on a forearm, a head, and a thigh, each of which represents the typical position of a smartwatch, an earbud, and a smartphone, respectively; the sampling rate of IMU sensing is 50 Hz.

\textit{Keyword spotting: } We use the keyword detection architecture proposed in~\cite{warden2018speech}. It takes a two-dimensional tensor extracted from the 1-second-long audio recording (time frames on one axis and MFCC on the other axis) as an input. The architecture consists of two convolutional layers, a global average pooling layer, and a fully-connected layer. For training and testing, we use the \emph{Keyword} dataset~\cite{mathur2019mic2mic} which consists of 65,000 speech keywords re-recorded at 16 kHz on three different embedded microphones (Matrix Voice, ReSpeaker, USB microphone) simultaneously; each file has 1-second long spoken keyword which belongs to one of 31 keyword classes.

We split all datasets to ensure independence between the training and test set. For the additional details, refer to~\cite{peng2018aroma, warden2018speech} for the architecture of sensing models and \cite{sztyler2016body, mathur2019mic2mic} for the datasets.

\textbf{Sensor workloads: } We consider two types of sensor workloads, \textit{static} and \textit{dynamic}. In the static workload, all devices are available all the time. It is used to investigate the overall performance of SensiX and the baselines. The dynamic workload is used to study the robustness of sensing in dynamic situations where some devices become temporarily unavailable, e.g., the battery runs out, or a user leaves a watch on the desk. We generate the four dynamic workloads with the following \textit{availability probability} values, 0.7, 0.8, 0.9, 1.0, respectively. For each workload, the availability of each device (either available or unavailable) is randomly decided based on the probability value. The decision is made independently to other devices.

\textbf{Comparison:} SensiX offers high-accuracy, robust sensing in the runtime environment with two main operations, device-to-device data translation and quality-aware device selection. To identify the impact of each operation, we considered the following baselines which also include variations of SensiX itself:

\begin{itemize}[leftmargin=*]
\item Single-avg: The traditional practice in context monitoring is to use a single, fixed device for a sensing model, e.g., either a smartphone, a smartwatch, or an earbud at runtime as shown in Figure~\ref{fig:overview} (a). Thus, each sensing task, we can consider three cases using different devices. Single-avg reports the average performance of these three cases.

\item SensiX-native: SensiX-native is built on top of the device discovery, but does not have the capability of the device-to-device data translation and quality-aware selection. Note that, by bringing AI execution to the system layer, SensiX can perform device discovery in the background, e.g., using Bluetooth and Wi-Fi scanning, and dynamically map sensing models to available devices. SensiX-native selects the device for model processing in a round-robin manner out of available devices.

\item SensiX-trans: SensiX-trans adopts the translation operation on top of SensiX-native, but does not have the quality-aware selection. That is, the device for model processing is selected in a round-robin manner, but the translation operation is added and used before the model execution if the given model was not trained in the selected device.

\item SensiX-QS: SensiX-QS selects the device with the quality-aware selection mechanism, but does not have the device-to-device data translation. 

\end{itemize} 


\textbf{Model and translation configuration: } We demonstrate the capabilities of SensiX in diverse training configurations where pre-trained models are available only from a subset of the devices. For the HAR task, we assume that two model instances (trained with data from \textit{head} and \textit{forearm} devices, respectively) are given and SensiX trains one translation function ($\textit{thigh} \rightarrow{} \textit{head}$). Indeed, as there is no model trained for \emph{thigh}-worn inertial sensors, SensiX would need to first perform translation on the data collected from the thigh to make it resemble the data from a training device (e.g., \emph{head}), before passing this data to the \emph{head} model for inference. Similarly, for keyword spotting model, we assume that one model instance (trained with the data from Matrix Voice) is given and SensiX trains translation functions for the other two devices (ReSpeaker $\rightarrow$ Matrix Voice and USB microphone $\rightarrow$ Matrix Voice).

\textbf{Performance metrics:} As a key performance metric, we consider the runtime accuracy of sensing models in the multi-device setting and compare SensiX against the aforementioned baselines. Since our datasets are imbalanced, we use the micro-averaged $F_1$ score~\cite{van2013macro}. In the dynamic workloads, the execution of a model cannot be supported if the dedicated device in Single-avg is unavailable or all the devices are unavailable in the SensiX variations. To reflect such scenarios, we set $F_1$ score to 0 during those moments when sensing is not supported. 

To understand the system behaviour comprehensively, we also consider other resource metrics such as energy consumption and execution time. We report these results with the micro-benchmarks in \S\ref{subsec:micro_benchmark}.

\subsection{Effect of SensiX operations on Accuracy Improvement}~\label{subsec:accuracy}

\subsubsection{Overall performance}


\begin{figure}[t]
    \centering
    \includegraphics[width=0.95\columnwidth]{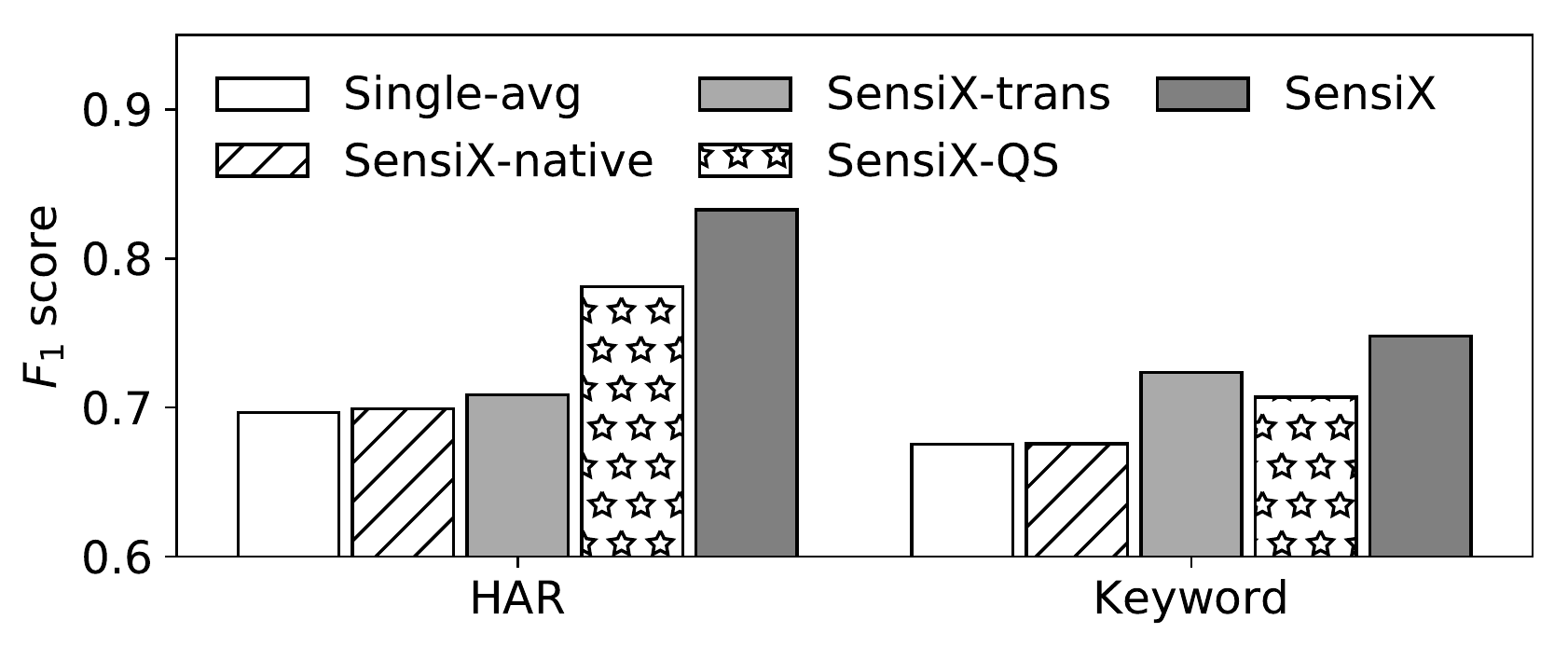}
    \caption{Overall Performance}
    \label{fig:overall_static}
\end{figure}

Figure~\ref{fig:overall_static} shows the overall performance under the static workload. The results show that SensiX increases the overall average $F_1$ score of sensing models by up to 0.13 without modifying them. More specifically, for the HAR task, SensiX achieves 0.83 of $F_1$ score, whereas the average $F_1$ score of the cases when a single device is used without any SensiX operations (Single-avg) is 0.70. It shows that the device-to-device data translation and quality-aware selection mechanisms enable applications to have more accurate results. We break down the performance by looking into and comparing the results of SensiX variations. As expected, Sensis-native shows similar performance to Single-avg because SensiX-native selects each device in turn, thus its performance converges to the average performance of the cases when each device is used. In HAR task, the improvement from the device-to-device translation, i.e., SensiX-trans (0.71) compared to SensiX-native (0.70), is not meaningful. This is because the performance of the translated \textit{thigh} data with the \textit{head} model is still much lower than other two device cases. However, SensiX-QS shows 0.78 of $F_1$ score, 8\% higher than SensiX-native, which shows the effectiveness of the quality-aware selection when multiple IMU devices are used. Also, when the translation operation is used together with the quality-aware selection, SensiX further achieves 5\% higher $F_1$ score than SensiX-QS. This shows that, even though the overall performance of the translation operation (from thigh data to head data) is not meaningful, our selection mechanism well spots the moments when the translated thigh data outperforms the data from the other two devices, and contributes to achieve the higher overall performance.

As shown in Figure~\ref{fig:overall_static}, the keyword spotting model shows the similar trend. SensiX increases the overall average $F_1$ score by 7\%; the $F_1$ score of SensiX and Single-avg is 0.75 and 0.68, respectively. Note again that, in the keyword spotting task, only one model trained with the data from Matrix Voice was used and thus ReSpeaker and USB microphone suffer poor accuracy when the translation operation is not applied, due to the heterogeneity issue described in \S~\ref{subsec:sensix_translation}. Different from the HAR case, we can observe the significant contribution of the translation operation. By adopting the translation function only, SensiX-trans achieves 0.72 of $F_1$ score.

It is important to note that the main goal of our work is not to train the most accurate sensing models and achieve high accuracy of each model, but to show that SensiX operations can increase the accuracy of pre-trained models in a multi-device sensing system. Our results, i.e., the relative improvement Single-avg to SensiX, clearly demonstrate this capability of SensiX. 

\subsubsection{Robustness in dynamic workloads} 

\begin{figure}[t]
    \centering
    \includegraphics[width=0.95\columnwidth]{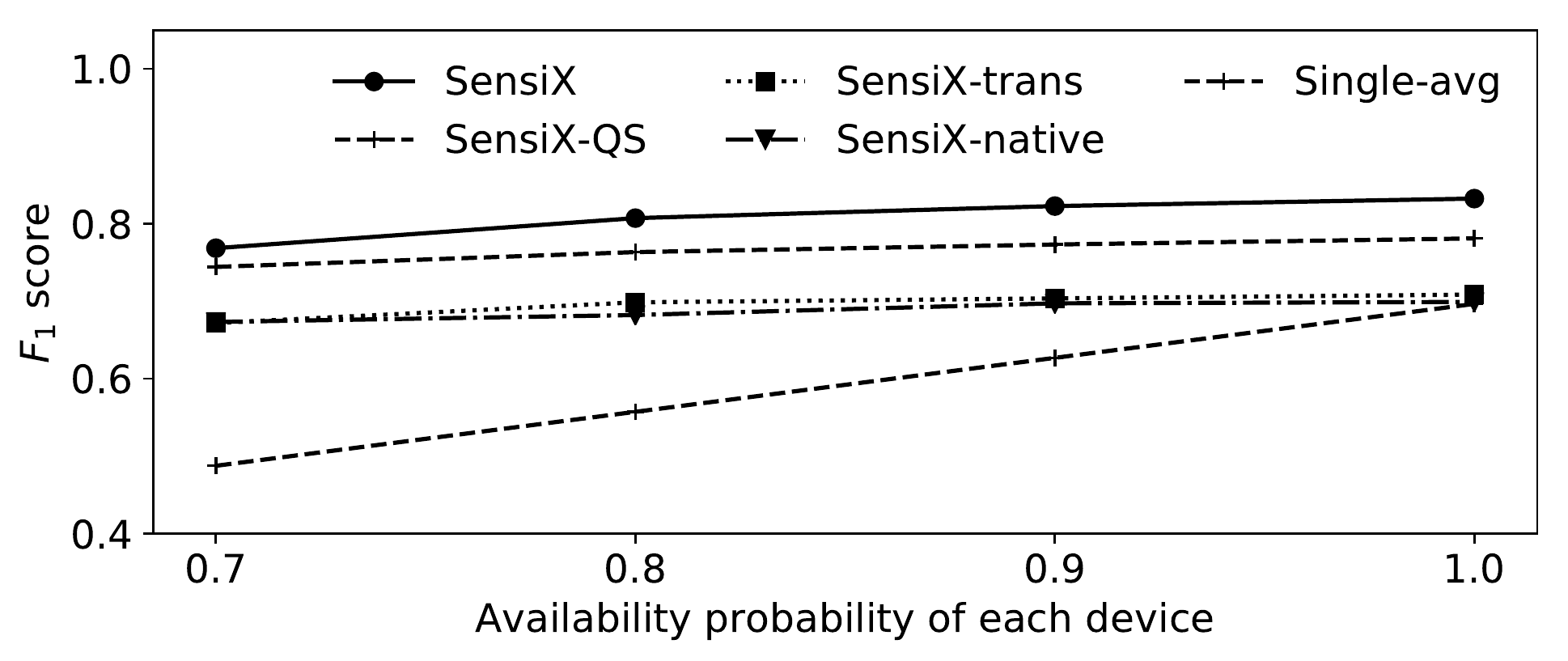}
    \caption{Performance of HAR in dynamic workloads}
    \label{fig:overall_realworld_dyanmic}
\end{figure}

\begin{figure}[t]
    \centering
    \includegraphics[width=0.95\columnwidth]{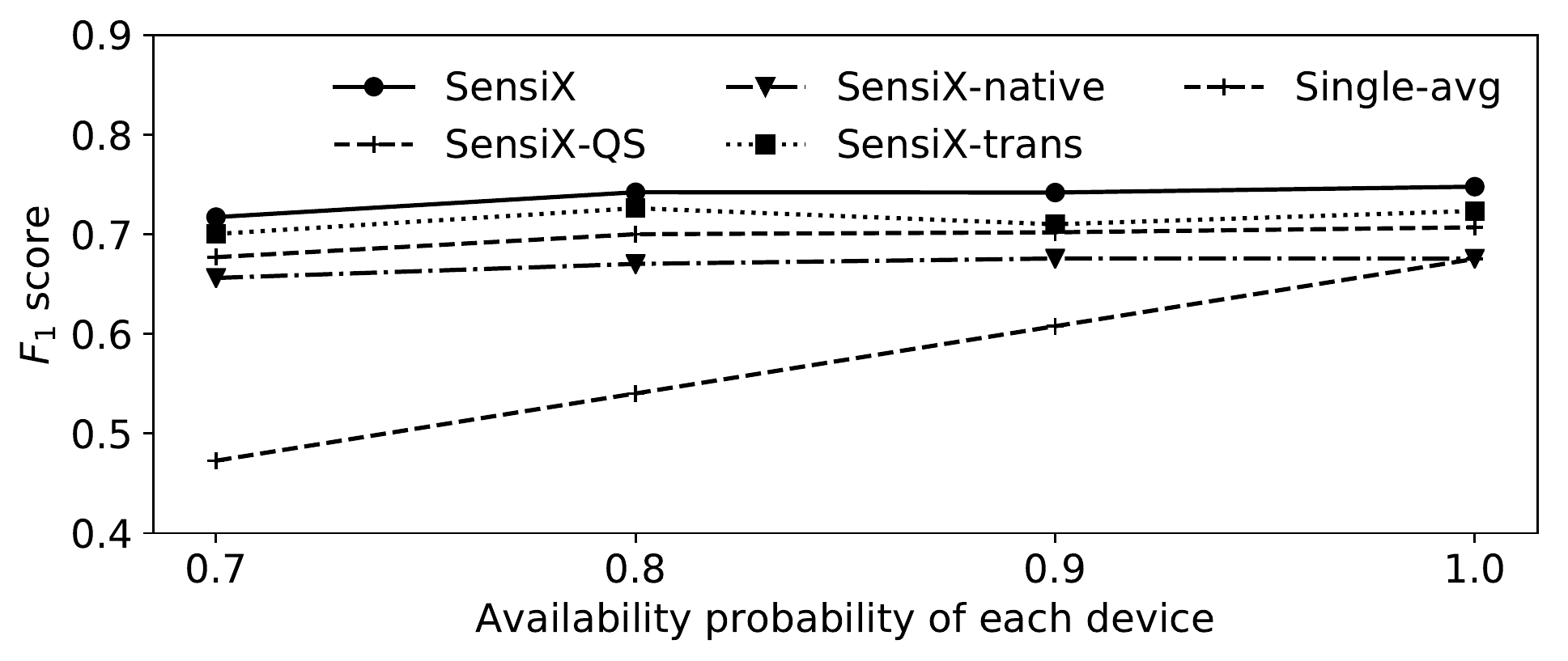}
    \caption{Performance of keyword in dynamic workloads}
    \label{fig:overall_audio_dyanmic}
\end{figure}

Figure~\ref{fig:overall_realworld_dyanmic} shows the $F_1$ score of the HAR model while increasing the availability probability from 0.7 to 1.0. When the probability is 0.7, each device is available with the probability of 0.7; for example, in this case, the probability of having two available devices out of three is 0.441 ($_3C_2 \times 0.7 \times 0.7 \times 0.3$). The results show that the SensiX capability with device discovery in multi-device environments achieves a higher level of robustness of sensing. Note again that, while dynamic mapping of available devices and sensing models are enabled by a unique feature of SensiX, separating execution operations from application space. As expected, Single-avg shows poorer performance as the availability probability becomes lower; it shows a linear relationship to the availability probability. However, the decrease of $F_1$ scores of SensiX variations is not significant. SensiX variations fail to deliver the model output only when all device are unavailable, which is very unlikely when there are multiple devices. For example, the $F_1$ score of SensiX is 0.77, 0.81, 0.83, and 0.83 when the availability probability is 0.7, 0.8, 0.9, and 1.0, respectively. 

Figure~\ref{fig:overall_audio_dyanmic} shows the result of the keyword spotting model in the dynamic workload. The results show the similar trend to the HAR model, but the decrease of $F_1$ score when the availability probability goes down from 1.0 to 0.7 is much smaller, e.g., $F_1$ score of SensiX is 0.75 and 0.72 with 1.0 and 0.7 of the probability, respectively. This is because the accuracy of keyword models is not much different across the devices when the translation operation is applied.

\subsection{In-depth Analysis}


\subsubsection{Device-to-device translation}

We look deeper into how much the translation operation improves the model performance.

\begin{figure}[t]
    \centering
    \includegraphics[width=1.0\columnwidth]{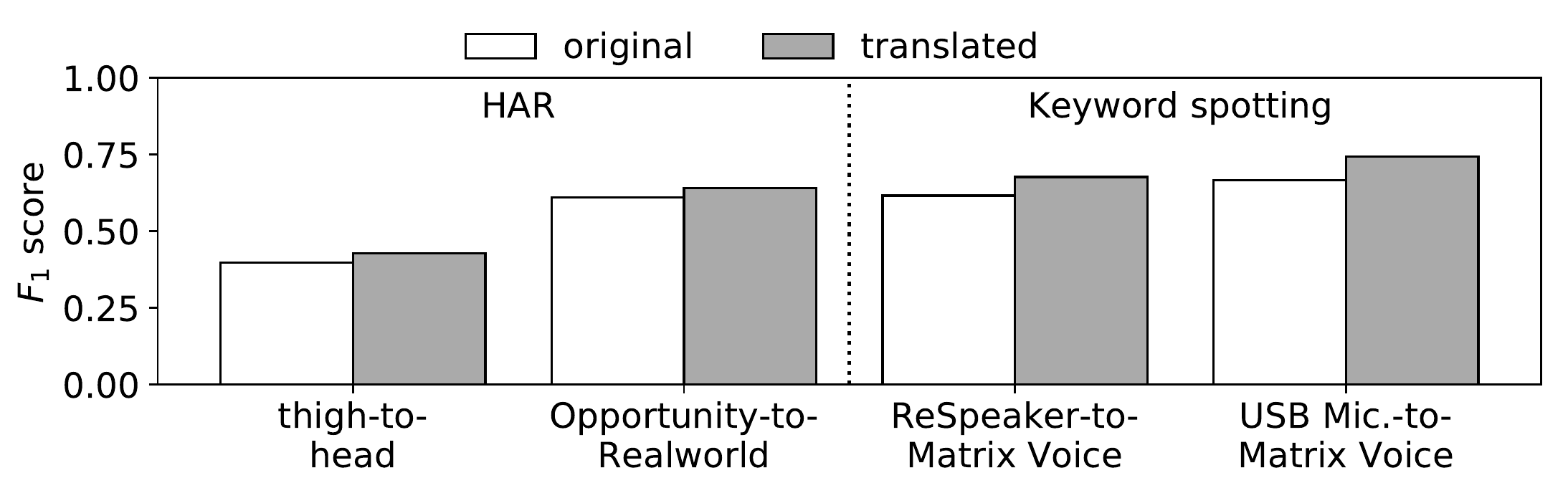}
    \caption{Effect of translation operation: (a) HAR (left) and (b) keyword spotting (right)}
    \label{fig:indepth_translation}
\end{figure}


\textbf{Translation between IMU sensors: } Here, we report two cases of translation between IMU sensors: (1) placement-to-placement translation, i.e., between the same type of IMU sensor, but with different placements and (2) device-to-device translation, i.e., between the different type of IMU sensor but with same placements. For the former, we report the performance of the \textit{thigh}-to-\textit{head} translation which was used in the experiments. For the latter, we introduce the opportunity dataset~\cite{roggen2010collecting} and investigate the performance of the translation from the \textit{left lower arm} device in Opportunity to the \textit{smartwatch on a forearm} in RealWorld, which can be seen to be placed in the same position. Since the collection configuration is different between Opportunity and RealWorld, we resample and normalise the Opportunity data before the translation; the sampling rate and accelerometer range are 30 Hz and $\pm3g$ for Opportunity and 50 HZ and $\pm2g$ for RealWorld. Figure~\ref{fig:indepth_translation} (a) shows the accuracy improvement when the translation operation is applied. More specifically, for the \textit{thigh} case, the performance of the \textit{original} and \textit{translated} is reported when the alternative model (here, \textit{head} model) is used with the original \textit{thigh} data and the translated \textit{thigh}-to-\textit{head} data, respectively. The results show that our translation operation improves the $F_1$ score from 0.39 and 0.43. In case of the translation of watch-worn devices from Opportunity to RealWorld, $F_1$ score of the original and translated is 0.61 and 0.64, when the model trained with the \textit{forearm} device in RealWorld is used with the resampled and normalised data from \textit{left lower arm} in Opportunity, without and with the translation, respectively.   

One may argue that, even with the accuracy improvement, translation between IMU sensors can be seen impractical due to their poor absolute accuracy. However, our experimental results in Figure~\ref{fig:overall_static} show that, together with the quality-aware selection, the translation provides the meaningful improvement of the system-wide accuracy. Considering that having device/placement-specific models requires significant burden for data collection and model engineering, SensiX can be used to complement applications when a new device is added, until the dedicated model is available. We believe we can further optimise the translation performance by adopting recent studies on domain adaptation for IMU sensors and motion models, e.g.,~\cite{10.1145/3380985,chen2019motiontransformer}.

\textbf{Translation between microphones: } Figure~\ref{fig:indepth_translation} (b) shows the results for the keyword spotting model. We assume that a model pre-trained on \textit{Matrix Voice} microphone is provided and is now tested with data from \emph{ReSpeaker} and \emph{USB Mic.}. As such, SensiX performs the following translation operations: \textit{ReSpeaker}-to-\textit{Matrix Voice} and \textit{USB Mic.}-to-\textit{Matrix Voice}. We observe that due to translation, the $F_1$ score of ReSpeaker increase from 0.62 to 0.68 (i.e., 6\% increase) and for USB Mic., the $F_1$ score increases from 0.67 to 0.74 (i.e., 7\% increase). These accuracy gains are significant for the Keyword Spotting model trained for Matrix Voice, whose best $F_1$ score is 0.77 when it is trained and tested on the same microphone. In other words, the translation operation is able to recover 40\% (for ReSpeaker) and 70\% (for USB Mic.) of the drop in $F_1$ score of the Keyword Spotting model due to microphone variability. 



\subsubsection{Quality-aware selection}

\begin{figure}[t]
    \centering
    \includegraphics[width=0.9\columnwidth]{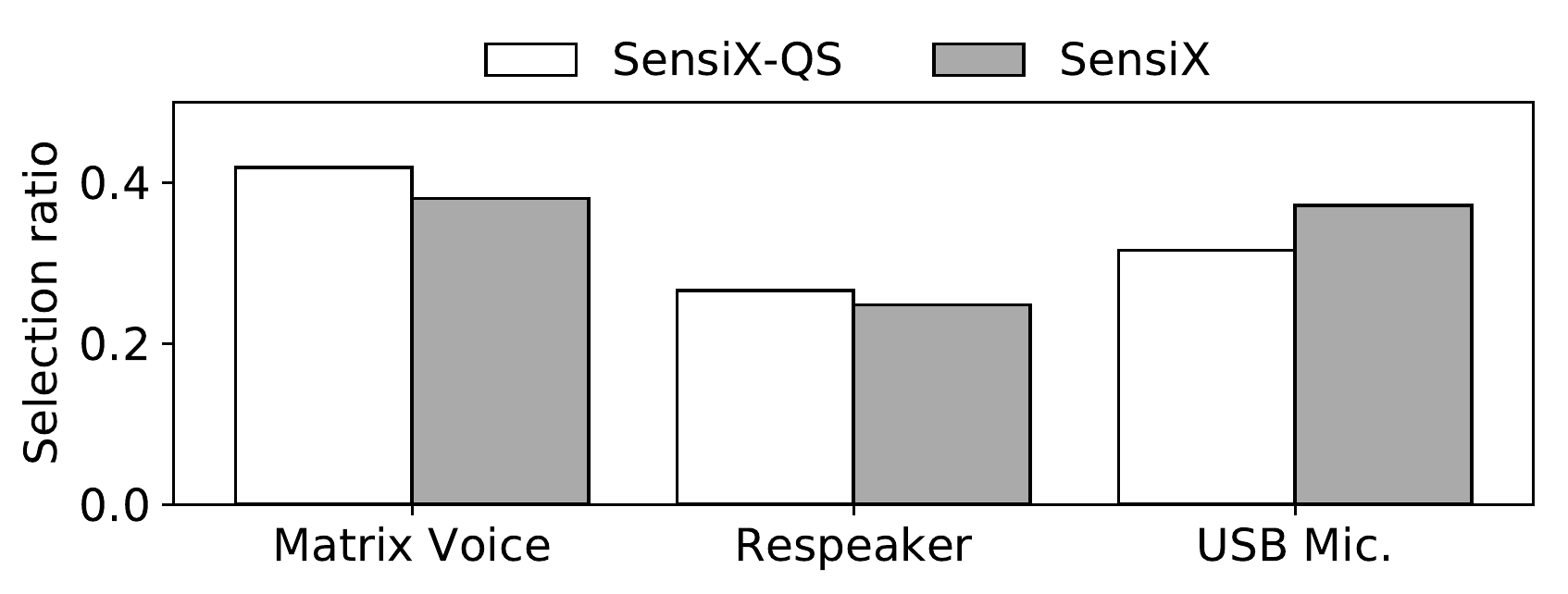}
    \caption{Ratio of selection (Keyword spotting)}
    \label{fig:indepth_selection}
\end{figure}

To have deeper understanding of the behaviour of the quality-aware selection, we look into how often each device is selected by the selection operator in the static workload. Figure~\ref{fig:indepth_selection} shows the selection ratio of three microphones on the keyword spotting model. The difference of ratio patterns between SensiX-QS and SensiX represents how the selection decision changes when the translation functions are added; note again that SensiX-QS does not have the translation operation. The results show that our selection mechanism well reflects the runtime quality, in two different ways. First, when the original signal is used without the translation (i.e., SensiX-QS), MatriX Voice is selected the most frequently, mainly due to its relatively higher performance. Second, on SensiX, the peformance of both ReSpeaker and USB Mic. improves from the translation to Matirx Voice as shown in Figure~\ref{fig:indepth_translation} (b). However, interestingly, while USB Mic. is selected more often, ReSpeaker is less selected. This is mainly because translated USB Mic. achieves comparable performance to Matrix Voice and thus produces more chances of being selected, especially instead of ReSpeaker.

\subsection{Micro Benchmark with SensiX Prototype}~\label{subsec:micro_benchmark}

\begin{figure}[t]
    \centering
    \includegraphics[width=1.0\columnwidth]{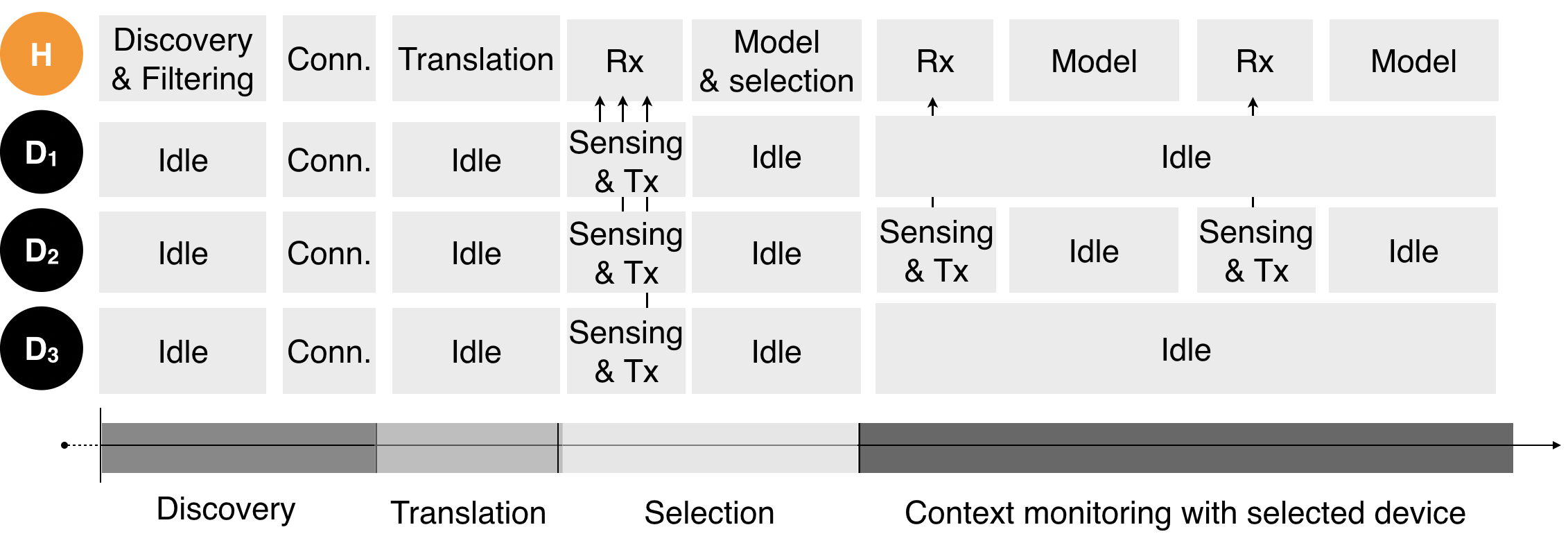}
    \caption{Main operations of SensiX; (H)ost device and sensor (D)evices. An example when $D_2$ is selected.}
    \label{fig:eval_operations}
\end{figure}

To understand the system behaviours, we conduct the micro-benchmark of the SensiX prototype with off-the-shelf devices as shown in Figure~\ref{fig:hardware}. Figure~\ref{fig:eval_operations} shows the main operations of SensiX. We first study the resource characteristics of model processing and then examine the system overhead of SensiX. We measure the energy cost using a Monsoon power monitor.

\begin{table}[t]
\small
\centering
\caption{Power cost for the motion model on sensor devices}
\label{tab:power_sensor}
\vspace{0.1in}
\begin{tabular}{cccc}
\textbf{Operation} & \multicolumn{3}{c}{\textbf{Power (mW)}} \\
 & \textbf{Pixel} & \textbf{LG Urbane 2} & \textbf{eSense} \\ \hline \hline
Idle & 28.1 & 27.8 & 6.6 \\
Sensing & 7.2 & 8.6 & \multirow{2}{*}{5.0 (BLE)} \\
Bluetooth Tx & 177.1 & 68.5 &  \\
\end{tabular}
\end{table}

\textbf{Model processing:} As described in \S\ref{subsec:implementation}, sensor devices act as a data source and model processing is conducted on a host device. Thus, the main status of the sensor device is either \textit{deactivated}, i.e., in idle mode, or \textit{activated}, i.e., sensing data and streaming it to the host device. Table~\ref{tab:power_sensor} shows the power profiles of sensor devices for the HAR task; we omit the result for the keyword spotting task due to the page limit. We report the power cost of sensing and BLE Tx for eSense together because its firmware does not support those functions separately. The transmission cost for Pixel 3 and LG Urbane is relatively much higher than that of eSense because Bluetooth classic is used for the communication with these Android devices. We expect that further energy saving can be achievable if the communication is developed on top of BLE. We leave it as future work.

\begin{table}[t]
\small
\centering
\caption{Resource cost for model processing on a host device}
\label{tab:model_cost}
\vspace{0.1in}
\begin{tabular}{cccc}
\textbf{Model} & \textbf{Parameters} & \textbf{Energy} & \textbf{Inference time}\\ \hline \hline
HAR & 385k & 0.86 mJ & 1.93 ms \\
Keyword & 1,846k & 28.12 mJ & 62.17 ms\\    
\end{tabular}
\end{table}

For model processing, the main operations of the host device are to receive sensor data and execute the model processing. Table~\ref{tab:model_cost} shows the energy consumption and inference time to be taken to process an instance of AI model execution on Raspberry Pi 3 and Coral USB accelerator. We observe that the execution of the keyword spotting model takes more energy and longer time due to its bigger size of the architecture. The average power to receive motion and audio data via Bluetooth classic remains around 10 mW.

\textbf{System overhead: } The major operations of SensiX beyond model processing are as follows. (1) SensiX discovers nearby devices by trying to establish the Bluetooth connection with paired devices, periodically. The interval of device discovery is configurable. The time to be taken to establish the connection differs depending on the type of sensor devices. It takes 0.9 sec and 5.7 sec for Android devices and eSense, respectively. We believe that the shorter time for Android devices comes from their optimisation of Bluetooth stack. (2) Once a new device is added and the corresponding model is not available, SensiX generates the translation function. Since training of translation function does not need the groundtruth data, a user just needs to wear the device for a period of time and SensiX performs training of the translation model locally or by offloading it to the third-party server. Once the translation model is available, SensiX processes it when the sensor data from the added device is needed. The average time for the translation model for one sample, i.e., one-second-long sensor data is around 20 ms and 480 ms for the HAR and keyword model, respectively. Their execution takes much longer than the model execution because the cycle GAN network in the translation model is much more complex. (3) SensiX performs the quality-aware selection at the interval of the duty cycle. For one selection operation, the overhead is to receive the sensor data and perform the sensing model for all devices during the assessment window (here, 1 sec.), i.e., additional 1.7 mJ and 28.30 mJ for the HAR and keyword spotting task, respectively; when three devices are available, additional processing of two more models is needed. Considering the selection interval, 10 seconds, the additional power overhead is 0.17 mW and 2.83 mW. (4) SensiX makes the execution schedule when multiple models are registered. Since the number of devices and models is relatively low in practice (around 5), the execution time and energy cost is negligible.

%% file: sections/discussion.tex
\section{Discussion}

\textbf{Why not multi-device fusion?} We assume that sensing models are built using sensor data from a \emph{single} sensory device. Recently, a number of studies on multi-sensory fusion have been conducted to maximise the inference accuracy while addressing potential system issues such as time synchronisation and missing data~\cite{ordonez2016deep,peng2018aroma,yao2017deepsense,yao2018qualitydeepsense,vaizman2018context}. While these works contributed substantially to achieve higher performance, we believe that they are not practical, yet to be used at the personal-edge in the multi-device environments. First, the fusion model requires all the devices involved in the training, to be activated all the time at runtime, thereby incurring significant system cost. Second, more importantly, considering the dynamics of multi-device environments, different fusion models are needed to be built and trained for all possible combinations of devices, which may not be feasible. For example, the fusion model trained with a smartphone and a smartwatch may be useless if a user forgets to wear the watch. Similarly, a new model will be needed if a user buys a new wearable device or an IoT device around a user becomes available.

\textbf{Beyond the accuracy: } SensiX can be easily extended to consider resource-related runtime metrics by adopting online profiling tools for energy~\cite{pathak2012energy} and transmission latency~\cite{gunther2005measuring} or by leveraging the benchmark study of the model performance, e.g.,~\cite{antonini2019resource,10.1145/3318216.3363299}. To this end, SensiX allows the policy to be specified as a cost function and selects the pipeline with the minimum cost output. For example, for the policy of minimising the total energy consumption, the corresponding cost function can be defined as \textit{f($D$, $M$) = total\_energy\_cost($D$, $M$)}, where $D$ and $M$ are a device and a sensing model, and total\_energy\_cost() is a function that returns the expected total energy consumption of devices when $D$ is selected for the processing of $M$. Several factors can be considered together by defining a cost function as their weighted sum.

\textbf{Generalisability of proposed techniques: } In this paper, we mainly focus on personal-edge environments with IMU and microphone sensors. Since device and data variabilities are common characteristics in multi-device environments, we believe that the main features of SensiX, a) separating execution of sensing models from application space and b) proposing device-to-device translation and quality-aware selection, are still valid in other edge environments with different type of sensors, especially, in case of multi-camera video analytics. However, translation and quality-assessment algorithms might be needed to be differently implemented depending on the environment and sensor type due to their different characteristics. We leave it as future work.




%% file: sections/related_work.tex
\section{Related Work}

\textbf{Context-aware middleware for BSNs:}
There have been extensive research efforts to develop context-aware middleware platforms for BSNs. They have developed abstractions to tackle challenges associated with context retrieval, device discovery, user mobility, and environmental changes, thereby making it easier to develop of context-aware applications. From a system's perspective, there have been two major directions in addressing multiple on-body sensors. The first direction is the dynamic sensor selection work~\cite{zappi2008activity,keally2011pbn,kang2008seemon,kang2010orchestrator}. Grounded on understanding of the effect of different characteristics on recognition accuracy, e.g., sensor type and composition, and device placement, they dynamically select the best sensor with the objective of optimising a system policy, e.g., maximising sensing accuracy, minimising energy cost. While they have presented execution strategies for various purposes, their consideration on the run-time accuracy has been limited. They mostly assumed that the run-time accuracy is same while the associated devices are available and made the decision based on the average accuracy. 

The second direction is to provide systematic support for building sensing applications in BSNs. They provide applications with high level of APIs and hide details of the system operations, e.g., resource management and coordination from multiple applications~\cite{kang2008seemon,kang2010orchestrator}, multi-sensor data fusion with inter-BSN communication, service discovery, ~\cite{fortino2015framework}, and application function virtualisation~\cite{kolamunna2016afv}.

Our work contributes to this rich body of BSN research by offering novel systematic aspects. First, SensiX automatically generates the sensing pipeline to make a given sensing model work in new, unseen devices by presenting device-to-device translation. Second, by adopting the selection mechanism~\cite{min2019closer}, SensiX further provides higher runtime accuracy.

\textbf{Offloading and partitioning of sensing pipelines: } One of the attempts to collaboratively use multiple devices in the mobile sensing area is code offloading and partitioning of sensing pipelines across sensors, a smartphone, an edge, and a cloud server~\cite{ha2014towards, cuervo2010maui, newton2009wishbone, rachuri2011sociablesense, kang2010orchestrator}. For a given sensing pipeline consisting of multiple sub-operations, they dynamically distribute the sub-operations over different devices in an optimal way by considering the application requirements and resource status on-the-fly. They make a distribution decision for a given pipeline without modifying it (by decomposing it into the sub-operations and distributing them). They mostly target the resource-related metric such as energy cost and latency and the decision does not affect the runtime accuracy. On the contrary, SensiX aims at achieving the highest runtime accuracy by actively optimising the runtime execution pipeline (the device-to-device translation and quality-aware selection), but without modifying the model itself. We believe SensiX can further optimise the resource use by decomposing the AI models into multiple sub-operations and adopting offloading and partitioning works.

\textbf{Streaming architectures: } From the point of view that sensor devices act as a data source, streaming processing architectures like Amazon Kinesis~\cite{kinesis} and Apache Kafka~\cite{kafka} can be seen similar to SensiX. Those systems are built as cloud-side messaging queue systems to deal with a high volume and number of streaming data durably, reliably, and with scalability. In this regard, they mainly focus on the data management layer, lacking the system aspects such as heterogeneity management for sensing quality. SensiX targets the execution pipeline layer to achieve high accurate and robust sensing on top of the data management layer.

%% file: sections/conclusion.tex
\vspace{-1mm}
\section{Conclusion}

We presented SensiX, a purpose-built runtime component for the personal edge to offer best-effort inference in a multi-device sensing environment. SensiX sits between the sensor devices and the corresponding sensory models in a personal edge device and comprises of two neural operators. A device-to-device data translation operator and a quality-aware device selection operator to cope with the device and data variabilities while externalising the model management and execution away from the applications. The combination of these two operators enables SensiX to boost runtime accuracy of sensory models in a multi-device sensory environment. We discussed different design cardinals, operational principles and implementation details of SensiX. We demonstrated the efficacy of SensiX through extensive testing with real-world motion and acoustic sensing workloads, including three difference public dataset and two different models. Our evaluation highlighted the ability of SensiX in boosting the overall runtime accuracy of different sensing models in multi-device sensing applications by 7-13\% and up to 30\% increase across different environment dynamics. We showed that this gain comes at the minimal expense of 3mW on the personal edge device, however with a significant reduction of development complexity and cost.
 
In the current version of SensiX, we assume that application developers provide model binaries. However, we envision that public repositories for pre-trained sensing models such as \cite{tensorflow_hub, pytorch_hub} can be easily used with SensiX, thereby allowing developers to choose and execute pre-trained models efficiently. We anticipate such ability of SensiX will enable application developers to focus on the application logic and model developers to focus on accurate model design.

In our future avenue of work, we plan to deploy and evaluate SensiX in real-world personal edge environment with multiple sensory devices. Besides, we want to explore the applicability of SensiX with other modalities, and in particular for vision-based applications. Finally, we aspire to assess the efficacy of SensiX in reducing code complexity by engaging developers in building multi-device sensing systems.

